\documentclass[12pt]{article}
\usepackage[utf8]{inputenc}
\usepackage{amssymb}
\usepackage{amsfonts}
\usepackage{amsmath}
\usepackage{cite}
\usepackage{comment}
\usepackage{graphicx}
\usepackage{multicol}
\usepackage{multirow}
\usepackage{setspace}
\usepackage{lineno}
\usepackage{makecell}
\usepackage{natbib}
\usepackage{caption}
\usepackage{url}
\usepackage{color}
\usepackage{layouts}
\setcitestyle{authoryear}
\usepackage[vmargin=1in,hmargin=1in]{geometry}
\DeclareMathOperator{\Var}{Var}
\newtheorem{proposition}{Proposition}
\newtheorem{corollary}{Corollary}

\newgeometry{hmargin=1in}

\title{\large \textbf{Optimal Survival Analyses With Prevalent and Incident Patients}}
\author{\large Nicholas Hartman}

\date{\normalsize Department of Biostatistics, University of Michigan, Ann Arbor, MI, U.S.A. \\
Kidney Epidemiology and Cost Center, University of Michigan, Ann Arbor, MI, U.S.A.}

\begin{document}

\maketitle

\begin{abstract}
Period-prevalent cohorts are often used for their cost-saving potential in epidemiological studies of survival outcomes. Under this design, prevalent patients allow for evaluations of long-term survival outcomes without the need for long follow-up, whereas incident patients allow for evaluations of short-term survival outcomes without the issue of left-truncation. In most period-prevalent survival analyses from the existing literature, patients have been recruited to achieve an overall sample size, with little attention given to the relative frequencies of prevalent and incident patients and their statistical implications. Furthermore, there are no existing methods available to rigorously quantify the impact of these relative frequencies on estimation and inference and incorporate this information into study design strategies. To address these gaps, we develop an approach to identify the optimal mix of prevalent and incident patients that maximizes precision over the entire estimated survival curve, subject to a flexible weighting scheme. In addition, we prove that inference based on the weighted log-rank test or Cox proportional hazards model is most powerful with an entirely prevalent or incident cohort, and we derive theoretical formulas to determine the optimal choice. Simulations confirm the validity of the proposed optimization criteria and show that substantial efficiency gains can be achieved by recruiting the optimal mix of prevalent and incident patients. The proposed methods are applied to assess waitlist outcomes among kidney transplant candidates.

\vspace{14pt}

\noindent \textit{Keywords}: Cox Proportional Hazards Model; Epidemiology; Kaplan-Meier; Left Truncation; Study Design
\end{abstract}

\singlespacing

\clearpage

\section*{Acknowledgements}
This preprint has not undergone peer review or any post-submission improvements or corrections. The Version of Record of this article is published in {\it Lifetime Data Analysis}, and is available online at \url{https://doi.org/10.1007/s10985-024-09639-6}. \newline

\noindent The data reported here have been supplied by the United Network for Organ Sharing as the contractor for the Organ Procurement and Transplantation Network. The interpretation and reporting of these data are the responsibility of the author(s) and in no way should be seen as an official policy of or interpretation by the OPTN or the U.S. Government. 

\clearpage

\section{Introduction}

In epidemiological studies with survival outcomes, the period-prevalent cohort is designed to achieve precision in both short- and long-term survival probability estimates while substantially limiting the required investments in patient follow-up \citep{Vonesh}. Prevalent patients, who have previously experienced the initiating event (e.g., a disease diagnosis) and survived until the study start date, are followed from the beginning to the end of the active study period. Incident patients, who experience the initiating event during the active study period, are followed immediately from the time of the initiating event to the end of the study. By including both prevalent and incident patients in the sample, the period-prevalent cohort design allows one to estimate survival probabilities over long time windows after the initiating event, even if the active study period is relatively short \citep{Wolfson}; we describe this point further and provide a detailed illustration of the design in Section \ref{sec:cohort}.

Most period-prevalent survival analyses from the existing literature have been designed to achieve some overall target sample size, without any consideration of how the relative percentages of prevalent and incident patients in the cohort impact the precision of the survival probability estimates and the power of statistical tests \citep{Burns,DOPPS}. While prevalent patients accrue survival time before the start of the study, and allow the investigator to efficiently assess long-term mortality risks with little follow-up, they are also subject to left-truncation, where early deaths that occur shortly after the initiating event are difficult to observe \citep{Schisterman}. In contrast, incident patients do not suffer from the issue of left-truncation, and contain valuable information about short-term mortality risks, but they must be followed for very long time periods to assess long-term mortality risks. Thus, there is an apparent trade-off between recruiting more prevalent versus incident patients, and even studies that have the same overall sample size may achieve substantially different levels of precision and power. Furthermore, there are no existing statistical methods available to rigorously maximize the total benefit from both prevalent and incident patients in period-prevalent survival analyses.

To address these important considerations, we derive optimization criteria that identify the most efficient mix of prevalent and incident patients to include in the sample. For survival curve estimation, we develop an objective function that expresses the variance of the estimated survival function in terms of the incident patient proportion, allowing one to minimize the total amount of uncertainty across the entire survival curve while flexibly allowing for a weighting scheme to emphasize precision in short- or long-term survival probability estimates. In addition, we prove that while the precision of survival curve estimators can improve substantially from mixing prevalent and incident patients under certain conditions, hypothesis tests based on the weighted log-rank test or Cox proportional hazards (PH) model are most powerful with an entirely prevalent or incident cohort. From this result, we show that the optimal cohort among these two choices (entirely prevalent or incident) can be identified by a simple comparison of two theoretical quantities. 

We make the proposed methods accessible through a user-friendly web application, which allows investigators to determine the optimal cohort design for estimation and inference based on a small amount of input parameters. Simulations show that the proposed study design method provides substantial efficiency gains over naive designs, such as one with an even mix of prevalent and incident patients, and appropriately adapts to the user-preferred weighting scheme. We apply these methods to study waitlist outcomes among kidney transplant candidates and observe that the optimal design produces meaningful differences in the estimated survival curve and inference results compared to naive designs. 

\section{Methods}
\label{sec:methods}

\subsection{Period-Prevalent Cohort}
\label{sec:cohort}

\begin{figure}[h!]
\centering
\hspace{-35pt}
\includegraphics[width=0.95\linewidth]{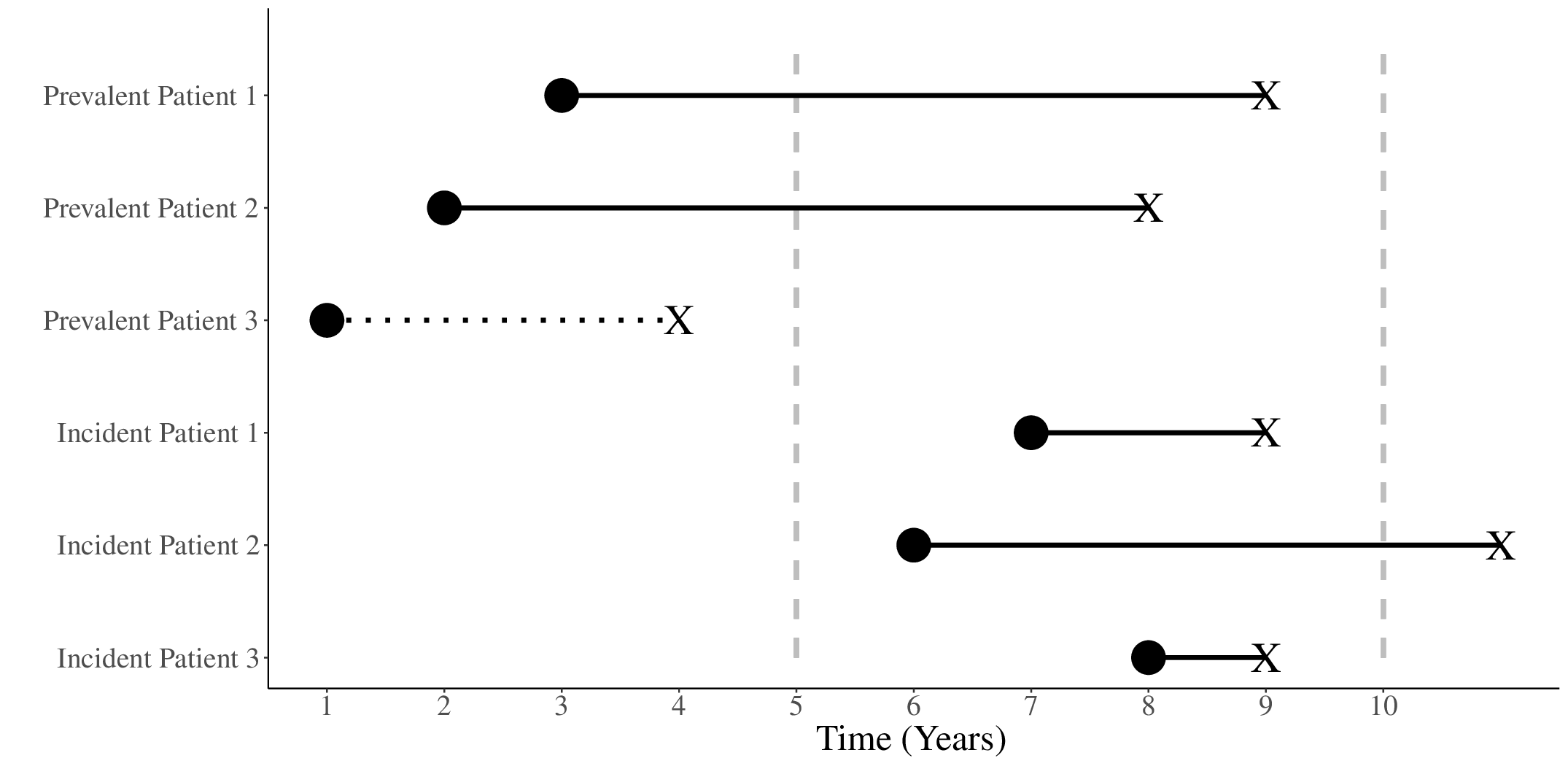}
\caption{An example period-prevalent cohort design. The solid circle represents the time origin (e.g., disease diagnosis) and the X symbol represents the endpoint of interest (e.g., mortality). Patients are recruited and followed during the active study window from years five through ten. Prevalent patients are recruited immediately at year five, whereas incident patients arrive throughout the active study window. Prevalent patients who experience the endpoint before the active study window (e.g., Prevalent Patient 3 with the dotted line) are unobserved and excluded from the sample. Patients who experience the endpoint after the active study window are right-censored at year ten (e.g., Incident Patient 2).}
\label{fig:design}
\end{figure}

Figure \ref{fig:design} demonstrates the period-prevalent study design. To make the exposition more concrete, we assume in this example that the outcome of interest is patient survival after the diagnosis of some disease. During the active study window, patients are followed and deaths are potentially observed. Prevalent patients, meaning those diagnosed with the disease before the active study window, must survive until the study start date to be included in the sample. As a consequence, prevalent patients who die very shortly after diagnosis are less likely to be included in the study; this corresponds to the well-known statistical issue of left-truncation \citep{Tsai,Kalbfleisch}. Incident patients, meaning those that are diagnosed within the active study window, are followed immediately from the time of diagnosis until the end of the study. However, unless the active study window is very long, deaths from incident patients that occur very late after diagnosis are unlikely to be observed within the study period. Thus, incident patients are usually subject to higher rates of right-censoring at later timepoints. By including both incident and prevalent patients in the sample, the period-prevalent design aims to balance the strengths and weaknesses of these patient types in assessing short- and long-term survival \citep{Wolfson}.

\subsection{Notation and Assumptions}
\label{sec:notation}

Let $T^*_i$ be a continuous random variable representing the underlying survival time for the $i^{\textrm{th}}$ patient, with cumulative distribution function $F(t)$ and survival function $S(t)=1-F(t)$. For prevalent patients, an underlying study entry time is denoted by the random variable $A^*_i$, with cumulative distribution function $H(a^*)$ and density function $h(a^*)$, and only patients with $A_i^* \le T_i^*$ can be observed in the sample. For incident patients, $A_i^*=0$. Let $\theta$ be the length of the active study period, such that the observed prevalent patients are right-censored at time $t=A_i^*+\theta$ and the observed incident patients are right-censored at time $t=U_i^*\theta$, where $U_i^*, 0 \le U_i^* \le 1$ is a random variable with cumulative distribution function $G(u^*)$, describing the proportions of the active study period that the incident patients are followed for. To simplify the description of the proposed methods, we assume for now that $A_i^*$ and $U_i^*$ are independent of $T_i^*$, and that no patients are lost to follow-up before the end of the active study period. In Section \ref{sec:nonadmin}, we describe extensions that account for non-administrative right-censoring. The observed survival time, $T_i$, and event indicator, $\delta_i$, are defined as  \begin{singlespace}
\begin{equation*} T_i=\begin{cases}\min(T_i^*,A_i^*+\theta), & \text{for prevalent patients} \\
\min(T_i^*,U_i^*\theta), & \text{for incident patients} \end{cases}\end{equation*}
\end{singlespace}
\noindent and \begin{singlespace}
\begin{equation*}
\delta_i=\begin{cases}I(T_i^* < A_i^*+\theta), & \text{for prevalent patients} \\
I(T_i^* < U_i^*\theta), & \text{for incident patients} \end{cases}, \end{equation*}
\end{singlespace} 

\noindent \newline where $I(\cdot)$ is the indicator function.  Finally, let $A_i=A_i^*|A_i^* \le T_i^*$ be a random variable corresponding to the observed study entry times. For a random sample of total size $n$, including some mix of prevalent and incident patients, we collect data $\{A_i, T_i, \delta_i\}, i=1,\dots,n$ as independent observations. Throughout this paper, we assume that the survival function $S(t)$ is estimated over some finite assessment interval $t \in [0,\tau]$, using the Kaplan-Meier approach and its extensions for left-truncated data \citep{KM,Wang}. 

\subsection{Optimal Estimation}
\label{sec:fun}

\subsubsection{Objective Function}

To quantify the amount of uncertainty in the estimated survival function at a given timepoint $t$, conditional on a certain mix of prevalent and incident patients in the sample, we consider the asymptotic variance of the Kaplan-Meier estimator \citep{KM}:\begin{equation}\\Var[\widehat{S}(t);\pi_I] \approx S(t)^2 \int_0^t \frac{f(r)}{Y(r;\pi_I)S(r)} \, dr, \label{eq:Var} \end{equation}

\noindent where $f(t)=F'(t)$ is the probability density function of the underlying survival times, and $Y(t;\pi_I)$ is the expected number of patients at risk at time $t$ given the proportion of incident patients in the sample $\pi_I$. In Appendix A, we derive the contributions of the prevalent and incident patient mix to the variance function by decomposing $Y(t;\pi_I)$ into two separate functions, $Y_P(t;\pi_I)$ and $Y_I(t;\pi_I)$, which represent the expected number of prevalent and incident patients still at risk at time $t$: \begin{equation}Y_P(t;\pi_I)=n (1-\pi_I) S(t) \frac{H(t)-H(t-\theta)}{\int_0^{\tau} S(t) h(t) \,dt},
\label{eq:Y_P}
\end{equation} \begin{equation} Y_I(t;\pi_I)=n \pi_I S(t) (1-G(t/\theta)), 
\label{eq:Y_I}
\end{equation}
\noindent where $Y(t;\pi_I)=Y_P(t;\pi_I)+Y_I(t;\pi_I)$. 

Intuitively, each of the functions in Equations (\ref{eq:Y_P}) and (\ref{eq:Y_I}) is a product of four interpretable components: the total sample size ($n$), the proportion of sampled patients that is either incident or prevalent ($\pi_I$ or $\pi_P=1-\pi_I$), the probability of the patient surviving until time $t$, and the probability of the patient entering the study before time $t$ and being followed until time $t$. Incident patients are typically right-censored earlier than prevalent patients, and the decay in $1-G(t/\theta)$ may cause $Y_I(t;\pi_I)$ to be close to zero for large values of $t$. In contrast, prevalent patients enter the study after already surviving for some period, and their right-censoring probabilities are often smaller for larger values of $t$. However, the delayed study entries of the prevalent patients can cause $H(t)$ and $Y_P(t;\pi_I)$ to be smaller for smaller values of $t$. Thus, as described in Section \ref{sec:cohort}, predominantly prevalent cohorts tend to have more patients at risk during later timepoints, whereas predominantly incident cohorts tend to have more patients at risk during earlier timepoints. In Sections \ref{sec:estim_timepoint} and \ref{sec:estim_curve}, we use this decomposition to uncover the impact of the patient mix on estimation efficiency. 

\subsubsection{Estimation at a Single Timepoint}
\label{sec:estim_timepoint}

We first address the situation in which primary interest lies in precise survival probability estimation at a single timepoint $t$. This scenario often arises in epidemiological applications where it is clinically meaningful to report on the five-year or ten-year survival probabilities of patients with a certain disease, for example. In Proposition \ref{result:est_timepoint} below (proof in Appendix B), we identify conditions that define the existence and properties of an optimal cohort which includes both prevalent and incident patients and minimizes $\Var[\widehat{S}(t);\pi_I]$ at the fixed timepoint of interest. 

\begin{proposition}
    For a given timepoint $t$ and sample size $n$, an optimal $\pi_I \in (0,1)$ that minimizes the asymptotic variance of $\widehat{S}(t)$ must orthogonalize functions $\gamma(r)=\frac{f(r)}{Y(r;\pi_I)^2}$ and $\psi(r)=1-G(r/\theta)-\frac{(H(r)-H(r-\theta))}{\int_0^{\tau} S(r) h(r) \,dr}$ over $[0,t]$ such that 
    $$\langle \gamma,\psi\rangle=\int_0^{t} \gamma(r)\psi(r) \, dr=0.$$
    If the above orthogonality condition does not hold for any $\pi_I \in (0,1)$, then the optimal value of $\pi_I$ is either zero or one. 
    \label{result:est_timepoint}
\end{proposition}  

The mathematical result described above has a useful interpretation from a study design perspective. For example, $S(r)\psi(r)$ represents the difference in at-risk probabilities for incident and prevalent patients as a function of time, and since $f(r) > 0, S(r) >0$, and $Y(r) > 0$, a necessary (but not sufficient) condition for an optimal $\pi_I$ to exist within $(0,1)$ is that $S(r)\psi(r)=0$ for some $r \in (0,t)$. Thus, as one would expect, if the probability of an incident patient being at risk is always higher than the probability of a prevalent patient being at risk over the entire assessment interval, then there is no benefit to mixing incident and prevalent patients in the cohort. In fact, for there to be a benefit in mixing incident and prevalent patients, we require the stronger condition that, for some $\pi_I \in (0,1)$, the difference in the at-risk probabilities is orthogonal to the hazard function divided by the square of the total at-risk function. Numerical methods may be applied to minimize $\Var[\widehat{S}(t);\pi_I]$ with respect to $\pi_I$, and the performance of the optimal $\pi_I$, denoted as $\pi_I^{\textrm{opt}}$, may be compared with some other value of $\pi_I$ through the Asymptotic Relative Efficiency (ARE) at time $t$: \begin{equation}
    \textrm{ARE}(t, \pi_I^{\textrm{opt}},\pi_I)=\frac{\Var[\widehat{S}(t);\pi_I]}{\Var[\widehat{S}(t);\pi_I^{\textrm{opt}}]}. 
\label{eq:ARE1}
\end{equation}

\subsubsection{Estimation of the Entire Survival Curve}
\label{sec:estim_curve}

While optimizing the precision at a single timepoint may be useful for studies with very focused research questions, it is often the objective to precisely estimate an entire survival curve over the assessment interval. Furthermore, the investigator may have varying levels of interest in the survival probability estimates at different timepoints across the interval (e.g., short-term versus long-term survival). As a natural generalization of the approach outlined in Section \ref{sec:estim_timepoint}, we specify the following objective function to summarize the total variance of $\widehat{S}(t)$ in terms of the parameter of interest $\pi_I$, while allowing for a flexible weight function to describe the relative importance of precision over time: \begin{equation} K(\pi_I)=\int_0^{\tau} W(t) Var\left[\widehat{S}(t); \pi_I\right] \,dt, \label{eq:objective} \end{equation}
 
\noindent where $S(t)$, $G(t)$, and $H(t)$ are pre-specified for use in Equations (\ref{eq:Var})-(\ref{eq:Y_I}), and $W(t)$ is any weight function with $W(t) > 0,$ for all $t \in [0,\tau]$, and $\int_0^{\tau} W(t) \,dt=1$. In Proposition \ref{result:est_curve}, we develop analogous conditions as in Proposition \ref{result:est_timepoint} to guarantee the existence and properties of an optimal patient mix for estimation of the entire survival curve.
\begin{proposition}
    For a given assessment window $[0,\tau]$, sample size $n$, and weight function $W(t)$, an optimal $\pi_I \in (0,1)$ that minimizes the asymptotic variance of $\widehat{S}(t)$ must orthogonalize functions $W(t)$ and $V(t)=S(t)^2 \int_0^t \frac{f(r)}{Y(r)^2}\big\{1-G(r/\theta)-\frac{(H(r)-H(r-\theta))}{\int_0^{\tau} S(r) h(r) \,dr}\} \, dr$ over $[0,\tau]$ such that $$\langle W,V\rangle=\int_0^{\tau} W(t)V(t) \,dt=0.$$
    If the above orthogonality condition does not hold for any $\pi_I \in (0,1)$, then the optimal value of $\pi_I$ is either zero or one. Under a uniform weight scheme with $W(t)=1/\tau$, the objective function reduces to $$K(\pi_I)=\frac{1}{\tau} \int_0^{\tau} \frac{f(r)(\tau-r)}{Y(r)S(r)} \, dr,$$
    \noindent and the optimization conditions are equivalent to the fixed-time problem described in Proposition \ref{result:est_timepoint}, with time $t=\tau$ and the original $\gamma(r)=\frac{f(r)}{Y(r)^2}$ replaced with $\gamma(r)=(\tau-r)\frac{f(r)}{Y(r)^2}$.
    \label{result:est_curve}
\end{proposition}

As in Proposition \ref{result:est_timepoint}, it can be shown that a necessary (but not sufficient) condition for the orthogonality to hold in Proposition \ref{result:est_curve} is that $S(r)\psi(r)=0$ for some $r \in (0,\tau)$ (i.e., the difference in the at-risk probabilities for incident and prevalent patients is zero at some timepoint). The above results provide the opportunity to highlight an important point that is central to the objectives of this paper. That is, even with a uniform weight function that corresponds to equal interest in all parts of the survival curve, the optimal cohort may be very different from the naive ``Even Mix'' cohort with 50\% of each patient type. For example, consider an extreme case, where $S(r)\psi(r) > 0$ for all $r \in (0,\tau)$. Then, $\pi_I^{\textrm{opt}}=1$ by Proposition \ref{result:est_curve}, even though the weight function is uniform. Therefore, statistically-informed period-prevalent study designs can provide valuable information for improving estimation efficiency. 

As in Section \ref{sec:estim_timepoint}, the optimal proportion of incident patients to recruit into the cohort is found by numerically minimizing Equation (\ref{eq:objective}), with $\pi_I^{\textrm{opt}}=\underset{\pi_I}{\textrm{arg\,min }} K(\pi_I)$ and the optimal proportion of prevalent patients is defined as $1-\pi_I^{\textrm{opt}}$. The ARE of the optimal cohort design, compared to any other cohort design, may be assessed through the ratio of the corresponding objective function values: \begin{equation} \textrm{ARE}(\pi_I^{\textrm{opt}},\pi_I)=\frac{K(\pi_I)}{K(\pi_I^{\textrm{opt}})}. 
\label{eq:ARE}
\end{equation}

\subsubsection{Parameter Specification}
\label{sec:algorithm}

Similar to traditional sample size calculations for any prospective study, the proposed optimization methods for period-prevalent cohorts require pre-specification of certain underlying study parameters. We give the following recommendations for selecting realistic versions of the necessary components: 
\begin{enumerate}
\item $S(t)$: The underlying survival distribution can be specified parametrically, based on some assumed density function for the survival times, such as one corresponding to the Exponential, Weibull, or Lognormal distributions. Historical data and expert opinion may be useful for informing this decision. 
\item $H(a^*)$: The distribution of the underlying study-entry times for prevalent patients, $A_i^*$ can be specified directly. In many cases, the investigator may be able to collect a pilot group of observed prevalent patients before the start of the study, which could be used along with inverse-probability weighting to help define $H(a^*)$ \citep{Wang}.  
\item $G(u^*)$: In many applications, it may be appropriate to assume that the incident patients are diagnosed at uniform timepoints within the active study period. In these cases, $U_i^*$ can be specified as a $\textrm{Uniform(0,1)}$ random variable, with $G(u^*)$ as the corresponding cumulative distribution function. If more flexibility is needed to accommodate complex patterns in diagnosis times, the $\textrm{Beta}(\textrm{shape1},\textrm{shape2})$ density may be a suitable alternative for specifying the distribution of $U_i^*$. 
\item $\theta$: The length of the active study window is often chosen based on the practical limitations of the study, such as the amount of resources available to recruit and analyze patients. 
\item $W(t)$: The weight function describes the relative importance of precision in the survival estimates throughout the assessment period $[0,\tau]$. Weight functions that are larger at earlier timepoints place greater emphasis on precision in short-term survival estimates, whereas weight functions that are larger at later timepoints place greater emphasis on precision in long-term survival estimates. In practice, it may be useful to specify $W(t)$ as the density function of a four parameter Beta distribution, $\textrm{Beta}(\textrm{shape1},\textrm{shape2},0,\tau)$,
which has the same properties as the traditional $\textrm{Beta}(\textrm{shape1},\textrm{shape2})$ distribution, but with support over $[0,\tau]$. This strategy ensures that $W(t)$ is a valid weight function (as described in Section \ref{sec:fun}), and it is extremely flexible, as the $\textrm{Beta}(\textrm{shape1},\textrm{shape2},0,\tau)$ density function can take on almost any shape. 

\end{enumerate}

\noindent After specifying the above parameter values, one may identify the most efficient value of $\pi_I$ by applying any function minimization algorithm. It is important to recognize that some parameter combinations can produce risk sets with an expected size of zero, causing the variance function to be infinite in Equation (\ref{eq:Var}). This numerical issue is usually an indication that $\tau$ is too large for the given patient mix, and the assessment interval should be narrowed.

\subsection{Non-Administrative Right-Censoring}
\label{sec:nonadmin}

For simplicity, we have assumed thus far that patients are only lost to follow-up due to administrative right-censoring from the study end date. However, other types of right-censoring are likely to occur in practice, such as patient drop-out before the end of the study. The theoretical results in Section \ref{sec:fun} can easily be extended to accommodate this non-administrative right-censoring by modifying $Y_P(t; \pi_I)$ and $Y_I(t; \pi_I)$ in Equations (\ref{eq:Y_P}) and (\ref{eq:Y_I}). To do this, one must first identify the most appropriate assumptions about the joint independence of $A_i^*$, $T_i^*$, and the non-administrative censoring variable, and whether it is possible for the non-administrative censoring event to occur before the study entry dates \citep{Qian}. In the simplest case with jointly independent non-administrative censoring, study entry, and failure, where the censoring event may occur before the study entry date, Equations (\ref{eq:Y_P}) and (\ref{eq:Y_I}) are multiplied by one additional term that describes the probability of the non-administrative censoring event occurring after time $t$. We present a real example of this data structure in our analyses of transplant candidates (Section \ref{sec:application}), where patients may be removed from the transplant waitlist at any time. 

\subsection{Optimal Inference}

\subsubsection{Two-Group Comparisons at a Fixed Timepoint}
\label{sec:inference_timepoint}

In some applications, one may perform statistical inference to assess whether the survival probability at a given timepoint is different across two independent groups (e.g., treatment versus control, exposed versus unexposed, etc.). Throughout this section and Section \ref{sec:inference_curve}, we use the subscripts ``1'' and ``2'' to denote mathematical quantities that are conditional on membership in either of the two groups. Using this notation, the null and one-sided alternative hypotheses for this fixed-time test are $H_0: S_1(t)=S_2(t)$ and $H_A: S_1(t) > S_2(t)$ respectively. 

The most commonly used test statistic for this type of comparison is the Z-score defined by $Z=(\widehat{S}_1(t)-\widehat{S}_2(t))\bigg/\sqrt{\Var[\widehat{S}_1(t)]+\Var[\widehat{S}_2(t)]}$. Under this testing framework, the expected value of $Z$ only depends on $\pi_I$ through the denominator. Therefore, assuming the same mix of prevalent and incident patients is used for both groups, the $\pi_I^{\textrm{opt}}$ value that maximizes the power of this test can be found by a straightforward extension of the methods in Section \ref{sec:estim_timepoint}, with $\Var[\widehat{S}_1(t); \pi_I]+\Var[\widehat{S}_2(t); \pi_I]$ as the objective function. Additional power can be achieved by allowing a different mix of prevalent and incident patients within each group and minimizing both $\Var[\widehat{S}_1(t); \pi_I]$ and $\Var[\widehat{S}_2(t); \pi_I]$ individually.

\subsubsection{Two-Group Comparisons of the Entire Survival Curve}
\label{sec:inference_curve}

In practice, two-group comparisons of patient survival are usually focused on differences in the entire survival curves, and one of the most popular inference methods for this type of analysis is the weighted log-rank test. The null and alternative hypotheses for this test are $H_0: S_1(t)=S_2(t)$ and $H_A: S_1(t)^{\phi(t)}=S_2(t)$ respectively, where $\log(\phi(t))=O(n^{-1/2}))$ is the log of the hazard ratio that we allow to depend on time. Under these hypotheses, \citet{Schoenfeld} proved that the optimally-weighted log-rank test uses weights that are proportional to $\log(\phi(t))$. In this section, we show that this optimally-weighted test can be optimized even further through statistically-informed sampling of prevalent and incident patients. 

As a first step, we decompose the probability of observing a failure at time $t$, denoted by function $D(t)$, into two separate components that reflect the separate contributions of prevalent and incident patients respectively
(proof in Appendix D): \begin{equation}
    D_P(t)=(1-\pi_I)f(t)\frac{H(t)-H(t-\theta)}{\int_0^{\tau} S(t)h(t) \,dt},
    \label{eq:D_P}
\end{equation}
\begin{equation}
    D_I(t)=\pi_If(t)(1-G(t/\theta)),
    \label{eq:D_I}
\end{equation}

\noindent where $D(t)=D_P(t)+D_I(t)$. Note that these formulas can be extended to accommodate non-administrative right-censoring using a similar approach as described in Section \ref{sec:nonadmin}. From Equations (\ref{eq:D_P}) and (\ref{eq:D_I}), we develop Proposition \ref{result:weighted_logrank} to describe the optimal period-prevalent cohort which maximizes the power of the weighted log-rank test.

\begin{proposition}
    For the optimally-weighted log-rank test with weights proportional to the log of the hazard ratio, $W(t) \propto \log(\phi(t))$ where $\log(\phi(t))=O(n^{-1/2})$, and under the alternative hypothesis $H_A: S_1(t)^{\phi(t)}=S_2(t)$, the power of the test is maximized when $\pi_I=1$ if
    $$\int_0^{\tau} \{\log{\phi(t)}\}^2 f(t)(1-G(t/\theta)) \,dt > \int_0^{\tau} \{\log{\phi(t)}\}^2 f(t)\frac{H(t)-H(t-\theta)}{\int_0^{\tau} S(t)h(t) \,dt} \,dt,$$
    and when $\pi_I=0$ otherwise. 
\label{result:weighted_logrank}
\end{proposition}

\noindent The above result, which suggests that there is no benefit to mixing prevalent and incident patients for the log-rank test, may at first appear somewhat paradoxical to the conclusions from Section \ref{sec:estim_curve}, which suggested that mixing prevalent and incident patients can improve the precision of the survival curve estimator. Mathematically, this difference occurs because the power of the log-rank test depends on a quantity that is a linear function of $\pi_I$, whereas the variance of the survival curve estimator is a convex function of $\pi_I$. 

\subsubsection{Multivariable Cox Proportional Hazards Models}
\label{sec:inference_cox}
Given the close relationship between the log-rank test and the Cox PH model, it is unsurprising that the results from Section \ref{sec:inference_curve} can easily be generalized to the multivariable regression case. Assume that the observed data are generated from an underlying Cox PH model, \begin{equation}\Lambda(t|\boldsymbol{X}_i)=\Lambda_0(t)\exp(\boldsymbol{X}_i^{\top}\boldsymbol{\beta}),
\label{eq:Cox}
\end{equation}

\noindent where $\Lambda_0(t)$ and $\Lambda(t|\boldsymbol{X}_i)$ denote the baseline and conditional cumulative hazard function, respectively, 
 and $\boldsymbol{X}_i$ is a vector of predictor variables, for the $i^{\textrm{th}}$ subject, with coefficient vector $\boldsymbol{\beta}$. Furthermore, without loss of generality, let $\beta_p$ denote the coefficient of primary interest, where the fitted model is used to assess the null hypothesis $H_0: \beta_p=0$.

\begin{corollary}
    For the multivariable Cox PH model in (\ref{eq:Cox}), with coefficient $\beta_p$ for the predictor variable of interest, and under the alternative hypothesis $H_A: \beta_p \ne 0$, the power of the score test is maximized when $\pi_I=1$ if
    $$\int_0^{\tau} f(t)(1-G(t/\theta)) \,dt> \int_0^{\tau} f(t) \frac{H(t)-H(t-\theta)}{\int_0^{\tau} S(t)h(t) \,dt}\,\,dt,$$
    and when $\pi_I=0$ otherwise. 
    \label{corollary:cox}
\end{corollary}

\noindent Note that Corollary \ref{corollary:cox} is equivalent to Proposition \ref{result:weighted_logrank} for the log-rank test with a fixed hazard ratio, and does not involve any properties of the predictor variables $\boldsymbol{X}_i$. Thus, any of the predictors, including the one that is of primary interest, can be continuous, discrete, or binary, and have any amount of correlation between them, and the result will still hold. 

\subsection{Web Application}

We make the proposed methods from this paper available through a Shiny web application ({\color{blue} \url{https://nh777.shinyapps.io/period_prevalent_cohort/}}) using the shiny package in R \citep{Shiny}. The user interface allows investigators to identify the optimal period-prevalent cohort and visualize the expected gains in precision and power from using this design. The application contains two separate pages for estimation and inference. In the upper-left panel of each page, the investigator specifies the study parameters as described in Section \ref{sec:algorithm}; various notes and definitions are available within a linked ``Study Parameter Guide'' to aid in making these choices. Immediately after specifying the parameters, the investigator can then visualize the corresponding shapes of $S(t)$, $H(t)$, and $W(t)$ in the upper-right area of the webpage to confirm that these functions are realistic. 

Once satisfied with the selections, the investigator may click the green ``Find Optimal Cohort'' button, which uses the input parameters to identify the optimal mix of prevalent and incident patients. On the ``Estimation'' page, this button triggers a quick computation (approximately one second) to minimize the objective function. A new display then appears in the lower-right panel of the user-interface, showing the optimal percentages of each patient type and a plot of the corresponding variance function over the assessment period. For comparison, the variance function is also shown for a naive ``Even Mix'' cohort, which always includes 50\% of each patient type regardless of the input parameters. On the ``Inference'' page, the ``Find Optimal Cohort'' button performs the comparison described in Corollary \ref{corollary:cox} and displays the optimal cohort (either 100\% incident or 100\% prevalent) with the expected number of failures and theoretical power.  

\section{Simulations}

\subsection{Optimal Estimation}
\label{sec:sim_estimation}

\subsubsection{Accuracy of the Objective Function}
\label{sec:accuracy}

We assess the performance of the proposed study design method through simulation. As a first evaluation, we verify that the derived theoretical formulas for $Y_P(t;\pi_I)$, $Y_I(t;\pi_I)$, and $\Var[\widehat{S}(t)]$ are accurate. On each iteration, we generate $T_i^*$ from an Exponential distribution with mean 10, and for prevalent patients, we generate $A_i^*$ from the same distribution, yielding a truncation probability (i.e., $P(A_i^* > T_i^*)$) of 50\%. We define  $\tau=10$, $\theta=7.5$, and $n=1000$. The theoretical quantities of interest are calculated from Equations (\ref{eq:Var})-(\ref{eq:Y_I}), and the empirical quantities are calculated from the Kaplan-Meier estimator (adjusting the risk sets for left-truncation) on the simulated data. The simulations are repeated for $\pi_I$=0.25, 0.50, and 0.75, each with 10,000 iterations. 

Figure S1 shows the resulting comparison between the theoretical and empirical versions of $Y_P(t;\pi_I)$, $Y_I(t;\pi_I)$, and $\Var[\widehat{S}(t)]$. We observe perfect agreement across all simulation settings, within a reasonable tolerance of Monte Carlo error, suggesting that the derived components of the objective function in Equation (\ref{eq:objective}) are valid. As expected, increasing $\pi_I$ reduces the variance at earlier timepoints and inflates the variance at later timepoints (Figure S1c), reflecting the inherent trade-offs in prevalent and incident patient recruitment. 

\subsubsection{Efficiency Gains}
\label{sec:precision}

Using a similar simulation structure as in Section \ref{sec:accuracy}, we quantify the potential efficiency gains that can be achieved through the optimal period-prevalent cohort design. For various weight functions $W(t)$, we minimize the objective function from Equation (\ref{eq:objective}) to determine the optimal mix of prevalent and incident patients in the sample. Then, for each of these optimal cohorts, we compute the empirical variance function, along with the empirical ARE, relative to a naive cohort design with 50\% incident patients. 

\begin{figure}[h!]
\centering
\hspace{-35pt}
\includegraphics[width=\linewidth]{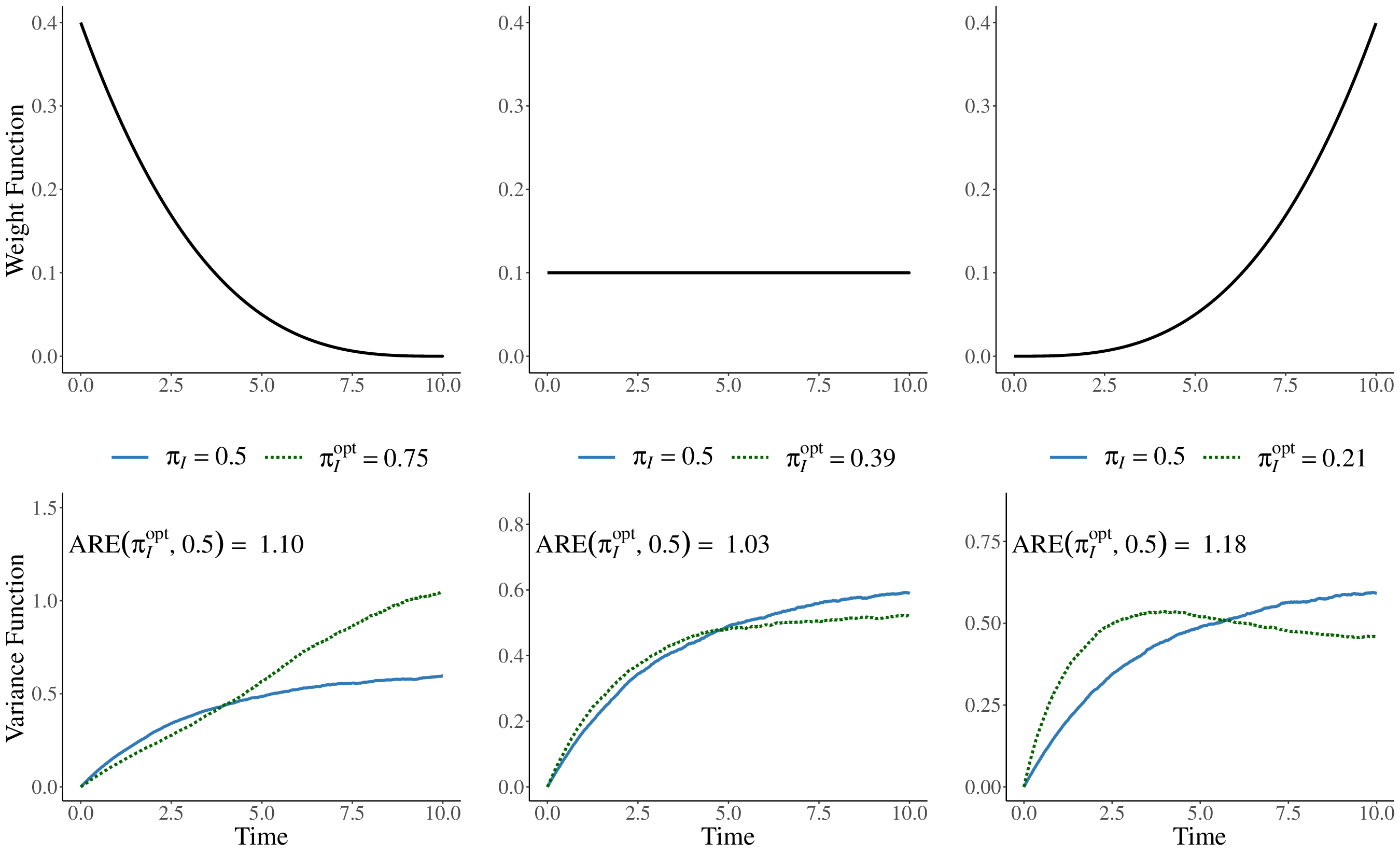}
\caption{(Top Row) Weight function and (Bottom Row) corresponding empirical variance function (multiplied by 1000) over the assessment interval $[0,\tau]$, for the ``even mix'' cohort, with $\pi_I=0.5$, and the optimally-efficient cohort, with the $\pi_I^{\textrm{opt}}$ that minimizes the weighted average of the theoretical variance function. The Asymptotic Relative Efficiency (ARE) of the survival estimates, comparing the optimally-efficient cohort to the even mix cohort, is shown for each choice of weight function. The empirical variance function is based on simulations with 10,000 iterations. }
\label{fig:precision}
\end{figure}

The numerical results in Figure \ref{fig:precision} show how the proposed study design method can adapt to the choice of weight function and identify the most efficient mix of prevalent and incident patients for each scenario. Using a right-skewed weight function (i.e., the $\textrm{Beta}(1,4,0,\tau)$ density function), we find that the optimal cohort contains 75\% incident patients, yielding a 10\% gain in efficiency over the naive cohort, which is mainly driven by a lower variance at earlier timepoints. In contrast, when a left-skewed weight function is used (i.e., the $\textrm{Beta}(4,1,0,\tau)$ density function), the optimal cohort contains just 21\% incident patients, yielding an 18\% gain in efficiency over the naive cohort and a substantially different variance function that decreases at later timepoints. Finally, using a uniform weight function, the optimal cohort contains 39\% incident patients, and has just a 3\% gain in efficiency over the naive cohort, reflecting their similarity under this setting. 

We note that the uniform weight function will not always correspond to an optimal cohort with a nearly even mix of prevalent and incident patients. In fact, when we repeat the above simulations for the uniform weight function, with different values of $\theta$, we observe in Table \ref{tab:ARE} that $\pi_I^{\textrm{opt}}$ can be very far from 0.5, and the length of the active study period has a major impact on the ARE. For short active study periods, $\pi_I^{\textrm{opt}}$ is much closer to zero because the incident patient censoring rate is high, and for long active study periods, $\pi_I^{\textrm{opt}}$ is much closer to one because the incident patient censoring rate is low. In both cases, there are substantial efficiency gains ($\approx$ 20-50\%) compared to the cohort with an even mix of prevalent and incident patients. Furthermore, we observe that cohort designs which contain only one type of patient (i.e., all prevalent or all incident) can be extremely inefficient for estimation, especially when the active study period is very short or long (Table \ref{tab:ARE}). 

\begin{table}[h!]
    \centering
    \caption{Length of the active study period ($\theta$), optimal proportion of incident patients ($\pi_I$), and empirical Asymptotic Relative Efficiency (ARE), comparing the optimal cohort to alternative cohorts with an even mix of prevalent and incident patients ($\pi_I=0.5)$, all incident patients ($\pi_I=1)$, and all prevalent patients ($\pi_I=0)$. The empirical ARE is based on 10,000 iterations, and an infinite empirical ARE occurs when some timepoints have zero patients at risk under the reference cohort. The length of the assessment period is $\tau=10$, and the uniform weight function is used in all settings.}
    \begin{tabular}{lllll}
    \hline\noalign{\smallskip}
    $\theta$ & $\pi_I^{\textrm{opt}}$ & $\textrm{ARE}(\pi_I^{\textrm{opt}},0.5)$ & $\textrm{ARE}(\pi_I^{\textrm{opt}},1)$ & $\textrm{ARE}(\pi_I^{\textrm{opt}},0)$  \\
    \noalign{\smallskip}\hline\noalign{\smallskip}
    1 & 0.09 & 1.56 & $\infty$ & 1.59 \\
    2.5 & 0.22 & 1.22 & $\infty$ & 2.37 \\
    5 & 0.39 & 1.03 & $\infty$ & 3.61 \\
    10 & 0.68 & 1.04 & 1.27 & 4.56 \\
    15 & 0.93 & 1.12 & 1.01 & 5.46 \\
    \noalign{\smallskip}\hline
    \end{tabular}
    \label{tab:ARE}
\end{table}

\subsection{Optimal Inference}

\subsection{Accuracy of the Optimality Criteria}
\label{sec:accuracy_inference}

We extend the simulation framework from Section \ref{sec:sim_estimation} to assess the proposed optimality criteria for inference with the log-rank test and Cox PH model. As a first step, we use the same simulation parameters as above, and compare the average numbers of prevalent and incident failures observed in the samples to the theoretical numbers obtained by integrating functions $D_P(t)$ and $D_I(t)$ (Equations (\ref{eq:D_P}) and (\ref{eq:D_I})) over $[0,\tau]$. Figure S2 shows perfect agreement between the empirical and theoretical results for a wide range of survival, arrival, and censoring distributions, confirming that the proposed derivation of the failure process, in terms of functions $S(t)$, $H(t)$, and $G(t)$, is suitable for defining the optimality criteria in Proposition \ref{result:weighted_logrank} and Corollary \ref{corollary:cox}. 

\subsection{Power Gains}
\label{sec:power_inference}

Having confirmed in Section \ref{sec:accuracy_inference} the theoretical validity of the proposed optimality criteria, we now demonstrate empirically the gains in power that are achieved by following the optimal design. For simplicity, we consider a two-sample test of the null hypothesis $H_0: S_1(t)=S_2(t)$, for all $t \in [0,\tau]$, as described in Section \ref{sec:inference_curve}. For each simulated dataset, we extract the P-value corresponding to the score test from the Cox PH model (which is equivalent to the log-rank test), and assess significance at the 5\% threshold. The survival data are simulated from (\ref{eq:Cox}), with constant baseline hazard function and coefficient $\beta=0.3$ to describe the effect of the group membership. We sample 500 patients from each group for a total sample size of $n=1000$, and vary the value of $\pi_I$ to compare the performances of different cohort designs. The censoring variable for incident patients is simulated as $U_i^*\theta$, with $U_i^*$ generated from a $\textrm{Beta}(c,1)$ distribution, where $c$ is a parameter that we vary such that the optimal cohort will eventually switch from 0\% to 100\% incident patients (as the number of observed failures among incident patients increases). 

\begin{figure}[h!]
\centering
\hspace{-35pt}
\includegraphics[width=\linewidth]{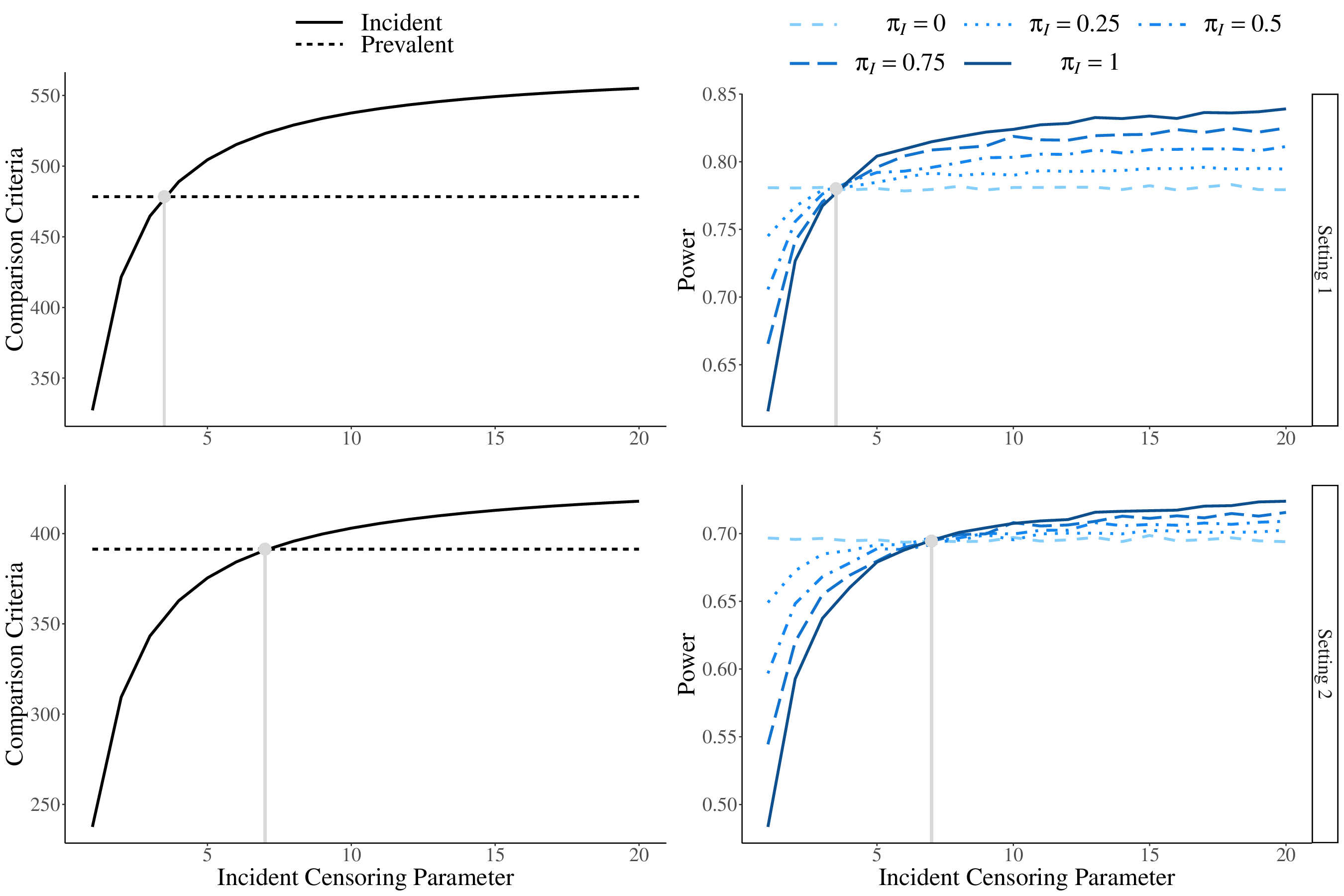}
\caption{(First Column) Theoretical comparison criteria from Proposition \ref{result:weighted_logrank} and Corollary \ref{corollary:cox}, for different distributions of the incident patient administrative right-censoring variable. If the value of the criterion for prevalent patients is larger than that for incident patients, then the entirely prevalent cohort is most powerful, and vice versa. (Second Column) Empirical power based on 100,000 simulation iterations, stratified by the proportion of patients that are incident ($\pi_I$). Simulation setting 1 (top row) sets $\theta=7.5$ and  simulation setting 2 (bottom row) sets $\theta=5$. The vertical light gray lines mark the points where the incident comparison criterion surpass the prevalent comparison criterion, or where the value of $\pi_I$ that yields the most power switches from zero to one.}
\label{fig:power_sim}
\end{figure}

The simulation results in Figure \ref{fig:power_sim} are consistent with the theoretical conclusions provided in Proposition \ref{result:weighted_logrank} and Corollary \ref{corollary:cox}. That is, the power of the test is a monotonic function of $\pi_I$ across all simulation settings, and there is no benefit to mixing prevalent and incident patients. Furthermore, the optimal cohort switches from 0\% to 100\% incident patients at exactly the first value of $c$ that makes $\int_0^{\tau} f(t)(1-G(t/\theta)) \,dt$ larger than $\int_0^{\tau} f(t) \frac{H(t)-H(t-\theta)}{\int_0^{\tau} S(t)h(t) \,dt}\,\,dt$. 

\section{An Application}
\label{sec:application}

We demonstrate the proposed study design method by performing a period-prevalent survival analysis of kidney transplant candidates. The Organ Procurement and Transplantation Network (OPTN) maintains a national registry of every patient in the United States (U.S.) who has ever been placed on the waitlist for a deceased donor kidney transplant. Before these candidates find a well-matched donor for transplantation, they rely on burdensome dialysis treatments that are associated with elevated risks of mortality \citep{Wolfe}. Thus, it is critical for these patients to get transplanted as quickly as possible, and for epidemiologists studying end-stage kidney disease, the time it takes for patients to receive transplants is an outcome of primary interest \citep{USRDS}.

Using the OPTN registry data, we explore the practical impact of our proposed design methods on transplant probability estimation and inference by constructing four hypothetical cohorts for comparison, each with a sample size of $n=5000$ but different mixes of prevalent and incident patients. In this example, we aim to precisely estimate the probability of receiving a transplant over time and perform inference to assess whether patient age is significantly associated with time to transplantation. The optimal cohort for estimation is identified by minimizing the objective function in Equation (\ref{eq:objective}), with a uniform weight function, and the optimal cohort for inference is identified based on the comparison criteria in Corollary \ref{corollary:cox} for the Cox PH model. 

We define the start and end dates of the active study period as 01/12/2021 and 01/12/2024, such that $\theta=3$, and set the upper bound of the assessment window to be $\tau=10$. Based on underlying registry data, we identify that a Weibull(shape=0.75, scale=4.25) distribution serves as a suitable model for the density of $T_i^*$, and a Weibull(shape=1.40, scale=4.25) distribution serves as a suitable model for the density of $A_i^*$. In addition, we model the density of $U_i^*$ with a $\textrm{Uniform}(0,1)$ distribution. Considering that patients can be removed from the waitlist at anytime and lost to follow-up for reasons other than the study concluding, we apply the extensions described in Section \ref{sec:nonadmin} to account for non-administrative censoring. 

Based on these parameter specifications and our proposed methodology, we find that the optimal cohort for survival curve estimation contains 74\% prevalent patients and 26\% incident patients. For comparison purposes, we also construct three other cohorts with $\pi_I=0.5$ (an even mix of prevalent and incident patients), $\pi_I=0$ (all prevalent patients), and $\pi_I=1$ (all incident patients). Furthermore, while the cohort with $\pi_I=0.26$ is considered to be optimal for estimation purposes, according to our proposed methods, the criteria from Corollary \ref{corollary:cox} suggest that the entirely prevalent cohort with $\pi_I=0$ is most powerful for inference on the relationship between patient age and time to transplantation (since $\int_0^{\tau} f(t)(1-G(t/\theta)) \,dt - \int_0^{\tau} f(t) \frac{H(t)-H(t-\theta)}{\int_0^{\tau} S(t)h(t) \,dt}\,\,dt=-0.08 < 0$ in this example). 

\begin{figure}[h!]
\centering
\hspace{-35pt}
\includegraphics[width=\linewidth]{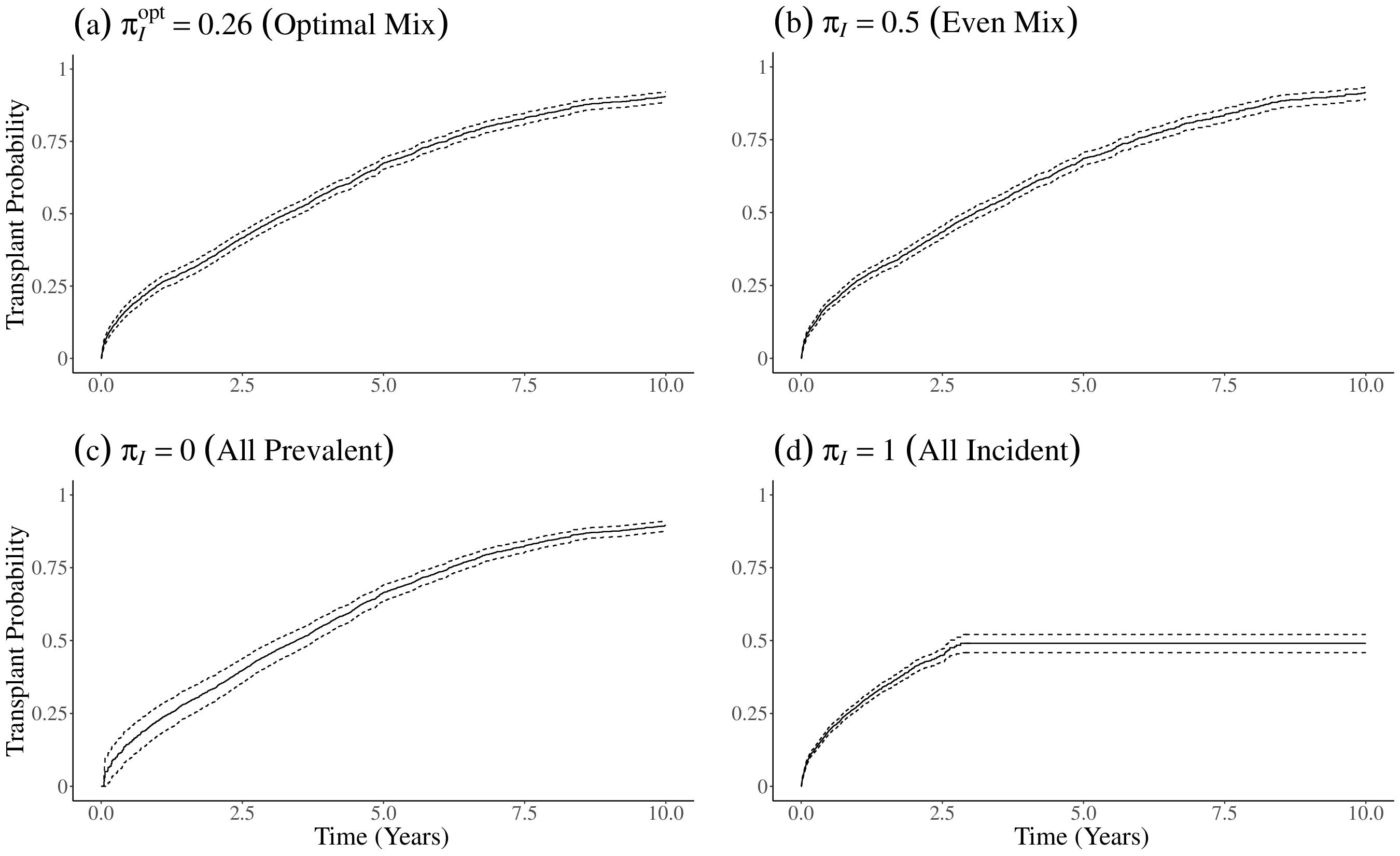}
\caption{Kaplan-Meier transplant probability curve estimates and 95\% confidence intervals for patients on the kidney transplant waitlist (sample size $n=5000$). The proportion of patients that are incident ($\pi_I$) is varied across panels to construct four hypothetical cohorts. The data for each cohort are sampled from the Organ Procurement and Transplantation Network (OPTN) database (as of January, 2024), using time period 01/12/2021-01/12/2024.}
\label{fig:KM}
\end{figure}

\begin{figure}[h!]
\centering
\hspace{-35pt}
\includegraphics[width=\linewidth]{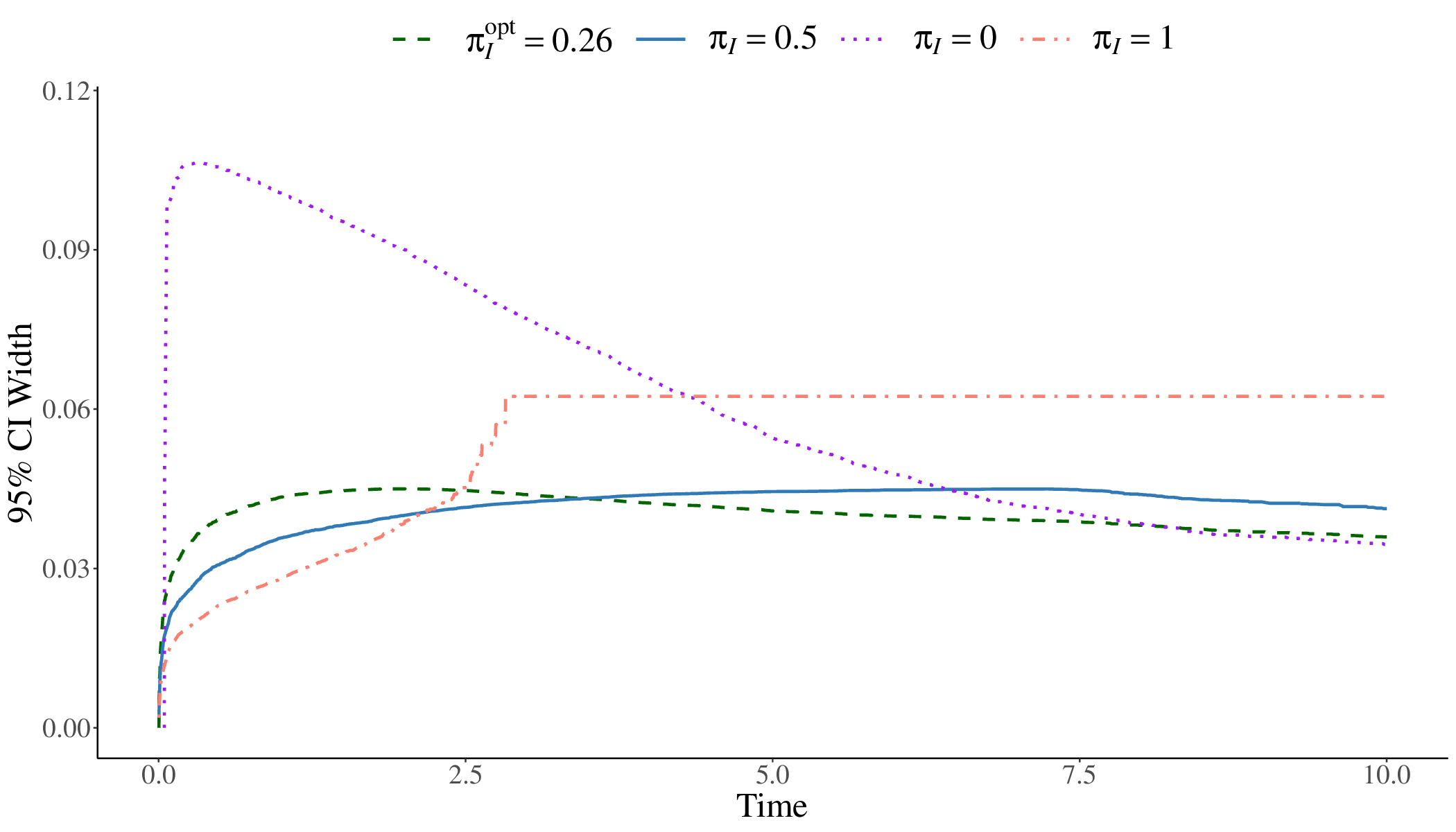}
\caption{Width of the 95\% confidence interval (CI) for the Kaplan-Meier survival estimator, stratified by the proportion of incident patients included in the sample. According to our proposed methods, the cohort with 26\% incident patients is considered optimal for survival curve estimation. The data for each cohort are sampled from the Organ Procurement and Transplantation Network (OPTN) database (as of January, 2024), using time period 01/12/2021-01/12/2024.}
\label{fig:KM_Interval}
\end{figure}

Figures \ref{fig:KM} and \ref{fig:KM_Interval} summarize the estimated transplant probability curves from the different study designs. Despite the overall sample size being fairly large in this study, we find clinically-meaningful differences in precision across the cohorts. For example, we observe that the entirely incident cohort (i.e., $\pi=1$) is a poor choice for estimation purposes, as no incident patients are at risk after $t=3$ and data are only available to estimate transplant probabilities for less than half of the assessment period. Thus, for $t > 3$, the estimated transplant probabilities from the incident cohort can only be extrapolated based on $\widehat{F}(3)$, which has a large variance. Similarly, the estimated transplant probabilities from the entirely prevalent cohort (i.e., $\pi=0$) have much wider confidence intervals at earlier timepoints compared to all other cohorts. In addition, Figure \ref{fig:KM_Interval} shows that the confidence intervals from the ``Even Mix'' cohort (i.e., $\pi=0.5$) are wider than those from the optimal cohort (i.e., $\pi=0.26$) for all $t \in (4,\tau]$, with only slightly narrower confidence intervals for $t \in [0,4]$. Thus, the optimal cohort achieves the best overall precision across the entire curve. 

As expected from the theoretical results of this paper, the observed performances of these cohort designs for statistical inference are very different from those for estimation. Table \ref{tab:real_power} shows the observed number of transplant events, the estimated hazard ratio for the age effect, and the corresponding score test statistic from the Cox PH model. Despite its relatively poor performance in precisely estimating the transplant probability curve, the entirely prevalent cohort has the most observed transplant events and the largest score test statistic, reflecting its expected optimality in terms of power as described by Corollary \ref{corollary:cox}. In contrast, the optimal cohort for estimation purposes and the ``Even Mix'' cohort, which produce the most precise transplant probability curves (Figure \ref{fig:KM}), have comparatively fewer observed transplant events and smaller score test statistics. Finally, the entirely incident cohort struggles the most to detect the age effect in this example (Table \ref{tab:real_power}), demonstrating the monoticity of the power function for the Cox PH model relative to $\pi_I$. 

\begin{table}[h!]
    \centering
    \caption{Number of observed transplant events, estimated hazard ratio, and score test statistic from a Cox PH model with  patient age as the predictor and time to transplant as the outcome. The proportion of patients that are incident ($\pi_I$) is varied to construct four hypothetical cohorts. According to our proposed methods, the cohort with $\pi_I=0.26$ is optimal for survival curve estimation, whereas the cohort with $\pi_I=0$ is optimal for inference. The data for each cohort are sampled from the Organ Procurement and Transplantation Network (OPTN) database (as of January, 2024), using time period 01/12/2021-01/12/2024. The score test statistic follows a Chi-squared distribution with one degree of freedom under the null hypothesis.}
    \begin{tabular}{llll}
    \hline\noalign{\smallskip}
    $\pi_I$ & Number of Transplant Events & Hazard Ratio & Score Test Statistic \\
    \noalign{\smallskip}\hline\noalign{\smallskip}
    0.26 & 1617 & 0.89  & 36.70 \\  
    0.5 & 1566 & 0.91  & 20.22 \\ 
    0 & 1727 & 0.88  & 44.21 \\ 
    1 & 1443 & 1.00  & \llap{$<$} 0.01 \\ 
    \noalign{\smallskip}\hline
    \end{tabular}
    \label{tab:real_power}
\end{table}

\section{Discussion}

In this paper, we have proposed an optimization method to identify the most efficient mix of prevalent and incident patients for estimation and inference in period-prevalent survival analyses. Through a user-specified weight function, the proposed optimizer for survival curve estimation adapts to the investigator's preferred focus on short- versus long-term survival assessments and appropriately balances the trade-offs in recruiting prevalent versus incident patients for these purposes. We have also shown analytically and empirically that, even with uniform weight functions, the optimal percentages of prevalent and incident patients may be far from an even mix with 50\% of each patient type, highlighting the need for informed patient recruitment strategies. In addition, we have proven that statistical inference based on the weighted log-rank test and Cox PH model is most powerful with an entirely prevalent or incident cohort, and there is not the same benefit from mixing prevalent and incident patients as there is with the survival curve estimator. 

The proposed study design method requires the investigator to specify certain underlying distributions (e.g., survival time, arrival time, censoring time) that are involved in the objective function and optimality criteria. While this requirement may be somewhat more challenging than traditional power analyses that only require the specification of fixed parameters, even cohorts based on inaccurate specifications may be substantially more efficient than those with entirely naive mixes of prevalent and incident patients. In fact, from Proposition \ref{result:weighted_logrank} and Corollary \ref{corollary:cox}, the most powerful cohort for inference (either 100\% prevalent or 100\% incident) is identified solely from a comparison of two mathematical quantites. Thus, if the underlying distributions are misspecified in a way that still preserves the ordering of these two quantities, the selected cohort will still be optimal.

The empirical results from this paper demonstrate that the mix of prevalent and incident patients, while often overlooked as a key study design consideration, can have a major impact on the results of period-prevalent survival analyses. Epidemiological studies have long relied on the period-prevalent cohort design to reduce the amount of time and effort required in assessing the effects of diseases \citep{Schisterman}. With the novel study design methods proposed in this paper, investigators have the potential to identify even better strategies for investing study resources, leading to reduced research costs and more reliable statistical analyses. 

\section*{Funding}

This work was supported in part by Health Resources and Services Administration contract HHSH250-2019-00001C. The content is the responsibility of the authors alone and does not necessarily reflect the views or policies of the Department of Health and Human Services, nor does mention of trade names, commercial products, or organizations imply endorsement by the U.S. Government.

\section*{Conflict of interest}

The authors declare that they have no conflict of interest.

\clearpage

\section*{\normalsize Appendix A: Derivation of $\boldsymbol{Y_P(t)}$ and $\boldsymbol{Y_I(t)}$}

For any given $t \in [0,\tau]$, let $R_i(t)$ denote a binary random variable that equals one if the $i^{\textrm{th}}$ observed patient from the sample is in the risk set at time $t$, and zero otherwise. Then, under the assumptions in Section 2.2, \begin{align*}
    E[R_i(t)]&=(1-\pi_I)P( T_i^* > t, t-\theta < A_i^* < t|A_i^* \le T_i^*,A_i^* < \tau)+\pi_I P(T_i^* > t, U_i^*\theta > t) \\
    &=(1-\pi_I)\frac{P(T_i^* > t,t-\theta < A_i^* < t, A_i^* \le T_i^*,A_i^* < \tau)}{P(A_i^* \le T_i^*,A_i^* < \tau)}+\pi_I P(T_i^* > t, U_i^*\theta > t) \\
    &=(1-\pi_I)\frac{P(T_i^* > t,t-\theta < A_i^* < t)}{P(A_i^* \le T_i^*,A_i^* < \tau)}+\pi_I P(T_i^* > t, U_i^*\theta > t) \\
    &=(1-\pi_I)\frac{P(T_i^* > t)P(t-\theta < A_i^* < t)}{P(A_i^* \le T_i^*,A_i^* < \tau)}+\pi_I P(T_i^* > t)P(U_i^*\theta > t) \\
    &=(1-\pi_I)S(t)\frac{H(t)-H(t-\theta)}{\int_0^{\tau} S(t) h(t) \, dt}+\pi_I S(t)(1-G(t/\theta)),
\end{align*}

\noindent where $P( T_i^* > t, t-\theta < A_i^* < t|A_i^* \le T_i^*,A_i^* < \tau)$ is the conditional expectation of $R_i(t)$ given the $i^{\textrm{th}}$ patient is prevalent, and $P(T_i^* > t, U_i^*\theta > t)$ is the conditional expectation of $R_i(t)$ given the $i^{\textrm{th}}$ patient is incident. It follows that 
    $Y(t; \pi_I)=\sum_{i=1}^n E[R_i(t)]= Y_P(t; \pi_I)+Y_I(t; \pi_I)$,
\noindent where $Y_P(t; \pi_I)=n(1-\pi_I)S(t)\frac{H(t)-H(t-\theta)}{\int_0^{\tau} S(t) h(t) \, dt}$ and $Y_I(t; \pi_I)=n\pi_IS(t)(1-G(t/\theta))$.

\clearpage

\section*{\normalsize Appendix B: Proof of Proposition 1}

\begin{align*}
    \frac{d}{d \pi_I}\Var[\widehat{S}(t); \pi_I]&=\frac{d}{d \pi_I} \int_0^{t} \frac{f(r)}{Y(r; \pi_I)S(r)} \, dr \\
    &=\int_0^{t} \frac{d}{d \pi_I} \frac{f(r)}{Y(r; \pi_I)S(r)} \, dr \\
    &=-\int_0^{t} \frac{f(r)}{Y(r; \pi_I)^2 S(r)} \frac{d}{d \pi_I} Y(r; \pi_I)\, dr \\
    &=-\int_0^{t} \frac{f(r)}{Y(r; \pi_I)^2 S(r)} nS(r)\bigg \{1-G(r/\theta)- \frac{H(r)-H(r-\theta)}{\int_0^{\tau} S(r) h(r) \, dr} \bigg \}\, dr \\
    &=-n \int_0^{t} \frac{f(r)}{Y(r; \pi_I)^2 } \bigg \{1-G(r/\theta)- \frac{H(r)-H(r-\theta)}{\int_0^{\tau} S(r) h(r) \, dr} \bigg \}\, dr,
\end{align*}

\noindent which equals zero when $\gamma(r)=f(r)/Y(r;\pi_I)^2$ and $\psi(r)=1-G(r/\theta)-\frac{H(r)-H(r-\theta)}{\int_0^{\tau} S(r) h(r) \, dr}$ are orthogonal over $[0,t]$. Furthermore, $$
    \frac{d^2}{d \pi_I^2}\Var[\widehat{S}(t); \pi_I]=2n \int_0^{t} \frac{f(r)}{Y(r; \pi_I)^3 } S(r) \bigg \{1-G(r/\theta)- \frac{H(r)-H(r-\theta)}{\int_0^{\tau} S(r) h(r) \, dr} \bigg \}^2\, dr > 0,
$$

\noindent since $f(r) > 0$, $S(r) > 0$, and $Y(r; \pi_I) > 0$ for all $r \in [0,t]$ and $\pi_I$. Thus, $\Var[\widehat{S}(t); \pi_I]$ is convex with respect to $\pi_I$, and the local minimizer defined by the orthogonality conditions in Result 1 is also the global minimizer. If a local minimizer does not exist, then the global minimizer must be either boundary value $\pi_I=0$ or $\pi_I=1$.

\clearpage

\section*{\normalsize Appendix C: Proof of Proposition 2}

\begin{align*}
    \frac{d}{d \pi_I}K(\pi_I)&=\frac{d}{d \pi_I}\int_0^{\tau} W(t) \Var[\widehat{S}(t); \pi_I] \, dt \\
    &=\int_0^{\tau} W(t) \frac{d}{d \pi_I}\Var[\widehat{S}(t); \pi_I] \, dt \\
    &=-n \int_0^{\tau} W(t) S(t)^2 \bigg\{ \int_0^{t} \frac{f(r)}{Y(r; \pi_I)^2 } \bigg \{1-G(r/\theta)- \frac{H(r)-H(r-\theta)}{\int_0^{\tau} S(r) h(r) \, dr} \bigg \}\, dr \bigg\} \, dt,
\end{align*}

\noindent which equals zero when $W(t)$ and $V(t)=S(t)^2\int_0^{t} \frac{f(r)}{Y(r; \pi_I)^2 } \bigg \{1-G(r/\theta)- \frac{H(r)-H(r-\theta)}{\int_0^{\tau} S(r) h(r) \, dr} \bigg \}\, dr$ are orthogonal over $[0,\tau]$. Furthermore, $$
    \frac{d^2}{d \pi_I^2}K(\pi_I)=2n S(t)^2\int_0^{\tau} W(t) \frac{d^2}{d \pi_I^2} \Var[\widehat{S}(t); \pi_I] \, dt  > 0,
$$

\noindent since $W(t) > 0$, $S(t) > 0$, and $\frac{d^2}{d \pi_I^2} \Var[\widehat{S}(t); \pi_I] > 0$ for all $t \in [0,\tau]$. Thus, by the same arguments in Appendix B, $K(\pi_I)$ is a convex function with respect to $\pi_I$, and the conditions in Result 2 for minimizing $K(\pi_I)$ must hold. 

\noindent Under the uniform weighting scheme $W(t)=1/\tau$, \begin{align*}
    K(\pi_I)&=\int_0^{\tau} \frac{1}{\tau} \Var[\widehat{S}(t); \pi_I] \, dt \\
    &=\frac{1}{\tau} \int_0^{\tau} \bigg\{ \int_0^{t} \frac{f(r)}{Y(r; \pi_I)S(r)} \, dr \bigg \} \, dt \\
    &=\frac{1}{\tau} \int_0^{\tau} \frac{f(r) (\tau-r)}{Y(r; \pi_I)S(r)} \, dr,
\end{align*}

\noindent which, by the results in Appendix B, is subject to the conditions in Result 1, replacing $\gamma(r)=f(r)/Y(r;\pi_I)^2$ with $\gamma(r)=f(r)(\tau-r)/Y(r;\pi_I)^2$. 

\clearpage

\section*{\normalsize Appendix D: Derivation of $\boldsymbol{D_P(t)}$ and $\boldsymbol{D_I(t)}$}

Using the results from Appendix A: \begin{align*}
    D(t)&=\lim_{\Delta \to 0} \frac{1}{\Delta} \bigg \{ (1-\pi_I)P(T_i^* > t-\Delta, t-\theta < A_i^* < t|A_i^* \le T_i^*,A_i^* < \tau)- \\ &\phantom{=n\lim_{\Delta \to 0}\frac{1}{\Delta}\bigg \{}(1-\pi_I)P(T_i^* > t, t-\theta < A_i^* < t|A_i^* \le T_i^*,A_i^* < \tau)+ \\ &\phantom{=n\lim_{\Delta \to 0}\frac{1}{\Delta}\bigg \{}\pi_I P(T_i^* > t-\Delta, U_i^*\theta > t) -\pi_I P(T_i^* > t, U_i^*\theta > t) \bigg \} \\
    &=\lim_{\Delta \to 0} \frac{1}{\Delta} \bigg\{(1-\pi_I)(S(t-\Delta)-S(t))\frac{H(t)-H(t-\theta)}{\int_0^{\tau} S(t) h(t) \, dt}+\\&\phantom{=n\lim_{\Delta \to 0}\frac{1}{\Delta}\bigg \{}\pi_I (S(t-\Delta)-S(t))(1-G(t/\theta))\bigg \} \\
    &=(1-\pi_I)f(t)\frac{H(t)-H(t-\theta)}{\int_0^{\tau} S(t) h(t) \, dt}+\pi_I f(t)(1-G(t/\theta) \\
    &=D_P(t)+D_I(t).
\end{align*}

\clearpage

\section*{\normalsize Appendix E: Proof of Proposition 3}

\citet{Schoenfeld} showed that the most powerful weighted log-rank test has weights proportional to the log of the hazard ratio, $\log(\phi(t))$, assuming that $\log(\phi(t))=O(n^{-1/2})$. If we denote $Z$ as the log-rank test statistic, then $Z \xrightarrow{d} \textrm{N}(\mu,1)$ under the alternative hypothesis, and the key result from \citet{Schoenfeld} was that  
$$\mu^2 \le n P_1 (1-P_1) \int_0^{\tau} \{\log(\phi(t)\}^2 D(t) \, dt,$$

\noindent where $P_1$ is the probability of being in the first comparison group, and equality only holds above under the optimal weighting scheme \citep{Xu}. Using this result, we further maximize the power of the optimally-weighted log-rank test by maximizing the upper bound above with respect to $\pi_I$.

\noindent The constants $n$ and $P_1$ do not depend on $\pi_I$ and can be ignored. Based on the derivations in Appendix D, \begin{align*}
    \int_0^{\tau} \{\log(\phi(t)\}^2 D(t) \, dt 
    &=\int_0^{\tau} \{\log(\phi(t)\}^2 (D_P(t)+D_I(t)) \, dt \\
    &=a+b\pi_I, 
\end{align*}

\noindent where $$a=\int_0^{\tau} \{\log(\phi(t)\}^2f(t)\frac{H(t)-H(t-\theta)}{\int_0^{\tau} S(t) h(t) \, dt} \, dt$$ \noindent and $$b=\int_0^{\tau} \{\log(\phi(t)\}^2f(t)\bigg\{1-G(t/\theta)-\frac{H(t)-H(t-\theta)}{\int_0^{\tau} S(t) h(t) \, dt}\bigg\} \, dt.$$ Thus, the upper bound of $\mu^2$ is a linear function of $\pi_I$, and the power of the optimally-weighted log-rank test is maximized when $\pi_I=0$ or $\pi_I=1$ if $b < 0$ or $b > 0$ respectively, proving the result.

\clearpage

\section*{\normalsize Appendix F: Proof of Corollary 1}

\citet{Hsieh} extended the power formula from \citet{Schoenfeld_Cox} for the Cox PH model to accommodate continuous predictor variables of interest with an arbitrary number of adjustment variables. Assuming a constant hazard ratio $\phi$, they showed that the score test statistic converges in distribution to $\textrm{N}(\mu,1)$ under the alternative hypothesis, where $\mu=\sqrt{n\{\log(\phi)\}^2 \sigma^2_X (1-\rho^2) \int_0^{\tau} D(t) \, dt}$, $\sigma^2_X$ is the variance of the predictor of interest, and $\rho^2$ is the coefficient of determination from regressing the predictor of interest onto the adjustment variables. Furthermore, maximizing the power of the score test is equivalent to maximizing $\mu^2$, which only depends on $\pi_I$ through $\int_0^{\tau} D(t) \, dt$. Following the same approach as in Appendix E:
$$\int_0^{\tau} D(t) \, dt =\int_0^{\tau} (D_P(t)+D_I(t)) \, dt=a+b\pi_I,$$ 
\noindent where 
$$a=\int_0^{\tau} f(t)\frac{H(t)-H(t-\theta)}{\int_0^{\tau} S(t) h(t) \, dt} \, dt$$ \noindent and $$b=\int_0^{\tau} f(t)\bigg\{1-G(t/\theta)-\frac{H(t)-H(t-\theta)}{\int_0^{\tau} S(t) h(t) \, dt}\bigg\} \, dt.$$

\noindent Thus, the power of the score test from the multivariable Cox PH model is maximized when $\pi_I=1$ or $\pi=0$ if $b < 0$ or $b > 0$, proving the corollary.

\clearpage

\setcounter{figure}{0}
\renewcommand{\thefigure}{S\arabic{figure}}

\begin{figure}[h!]
\centering
\hspace{-35pt}
\includegraphics[width=\linewidth]{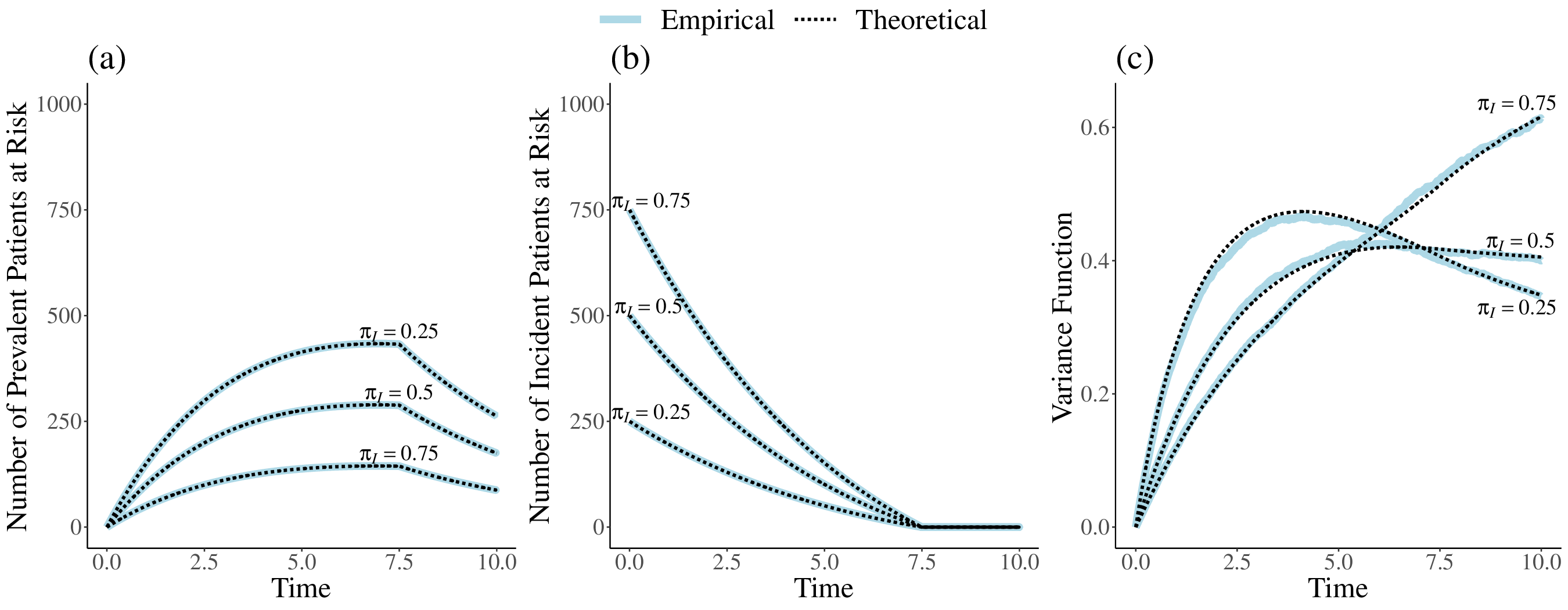}
\caption{A comparison of the empirical and theoretical quantities related to the precision of survival function estimators, for timepoints $t=0$ through $t=10$: (a) number of prevalent patients at risk, (b) number of incident patients at risk, and (c) variance function from Equation (1) (multiplied by 1000). The empirical quantities are generated from simulations based on 10,000 iterations, and the theoretical quantities are derived in Section 2. Curves are shown for different proportions of incident patients included in the sample (i.e., $\pi_I$=0.25, $\pi_I$=0.5, and $\pi_I$=0.75), with a total sample size of $n=1000$.}
\label{fig:accuracy}
\end{figure}

\clearpage

\begin{figure}[h!]
\centering
\hspace{-35pt}
\includegraphics[width=\linewidth]{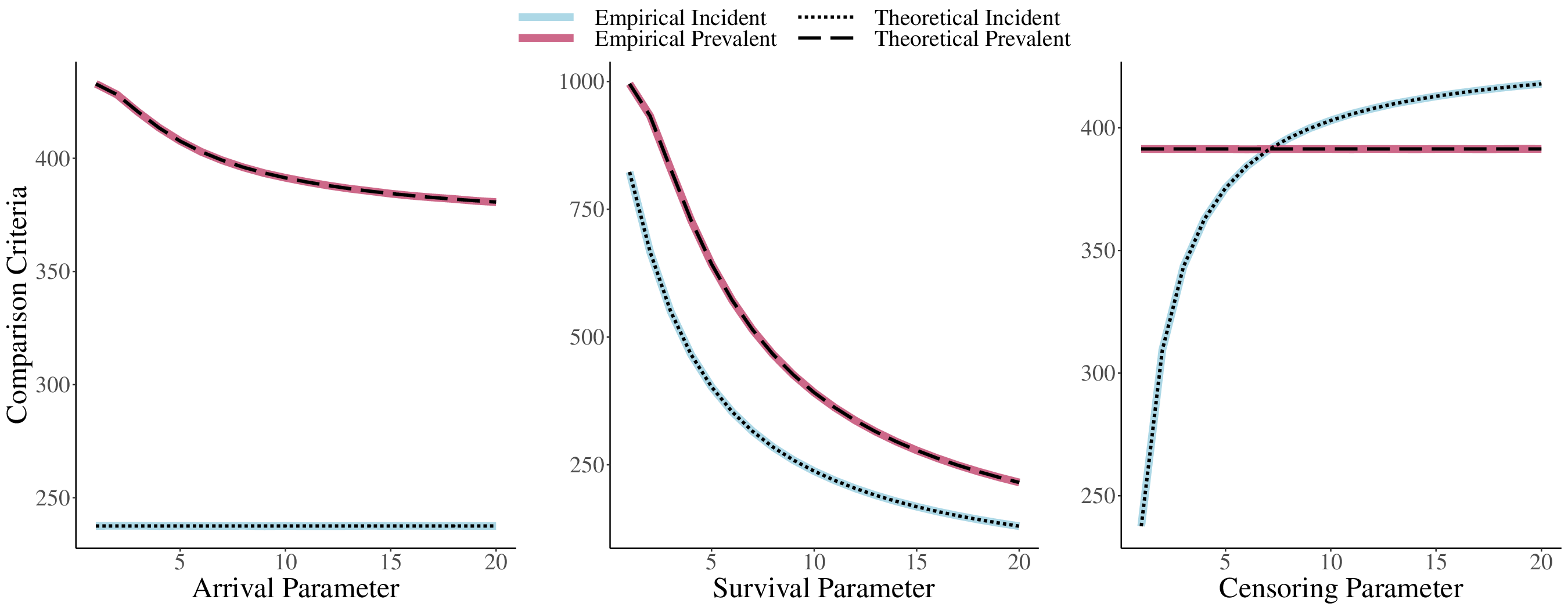}
\caption{A comparison of the empirical and theoretical criteria related to the power of the Cox PH model, for different arrival time, survival time, and censoring time distributions. The empirical quantities are generated from simulations based on 100,000 iterations, and the theoretical quantities are calculated from the formulas in Corollary 1. Curves are shown for prevalent and incident patient types.}
\label{fig:accuracy2}
\end{figure}
\clearpage

\bibliographystyle{apalike} 
\bibliography{main.bib}

\end{document}